\documentstyle[preprint,prc,aps,epsfig]{revtex}
\begin{document}

\title{Formal comparison of SUSY in the nuclear U(6/2) model and 
       in quantum field theory}

\author{R. V.~Jolos$^{1,2}$  and P.~von Brentano$^1$ , }

\address{$^1$ Institut f\"ur Kernphysik, Universit\"at zu K\"oln, K\"oln, 
              50937 K\"oln, Germany}
\address{$^2$ Bogoliubov Laboratory of Theoretical Physics, Joint Institute 
         for Nuclear Research, 141980 Dubna, Russia}

\date{\today}

\tighten

\maketitle

\begin{abstract}
A nuclear physics example of the U(6/2) 
supersymmetry group is considered. It is shown that this group contains
a supersymmetric subgroup with a structure similar to the SUSY
model of the quantum field theory (QFT). 
A comparison of two models helps to
clarify the relation between the supersymmetry schemes of QFT and of 
nuclear physics. 
Using this similarity
relation between the numbers of the bosonic and
fermionic states similar to the fundamental relation in  QFT is obtained. 
For those 
supermultiplets with at least two fermions the numbers of the bosonic and 
fermionic states are equal as in QFT.
\end{abstract}

\section{Introduction}

Supersymmetry is a very interesting subject which is widely discussed
in different subfields of physics. Supersymmetry has a long and 
interesting history. It was discovered in the particle physics, however,
the ideas of supersymmetry have found in nuclear structure physics 
a very extensive application.
Several good examples of an approximate dynamical supersymmetry have been 
found there. The investigations of the dynamical supersymmetries in
nuclear structure physics has been initiated by F.Iachello and the results
obtained are summarized in \cite{Iachello1}.

It was found that the combined algebras are particularly 
useful in a description
of the properties of odd--even nuclei. Moreover, an attempt to get a
classification in which several nuclei including even and odd ones are
described by the same Hamiltonian has been realised and very interesting 
results have been obtained \cite{Iachello1,Frank,Iachello2,Balantekin,Iachello3,Bijker1,Bijker2,Bijker3,Vervier,Vervier1,Cizewski,Warner,Vervier2}

However, the formalism to introduce the supersymmetries used is 
different in particle physics \cite{Wess,Fayet,Soff}
and in nuclear
structure physics. This makes
a comparison more difficult and also restricts the application 
of the examples found in nuclear physics.
In nuclear structure theory 
the main attention has been paid
to the construction of the Hamiltonians, which exhibit dynamical 
supersymmetries. Such Hamiltonians are represented by the sums of the
Casimir operators for an appropriate subgroup chain of U(6/$\Omega$),
where $\Omega$ is a maximum number of the fermions in a system. Taken
with the same parameters the Hamiltonians constructed in this way have
been applied to the description of different nuclei including even and odd ones.
However, previous works considered those subgroup chains in which 
already at the first step a classical Lie algebra formed by the operators
of the Bose sector of U(6/$\Omega$) appears.

For this reason,  the operators of the Fermi sector
(in the following called super operators), 
i.e. the operators transforming bosons
into fermions and vice versa \cite{Jolos}, do not appear explicitly 
in the subgroup chains.
The super operators of the basic U(6/$\Omega$) 
graded algebra
commute with a Hamiltonian which is a 
function of only the
Casimir operator of U(6/$\Omega$), which is the operator of the
total number of bosons plus fermions. 
Such a Hamiltonian is too simple to describe
real nuclei and it does not contain an interaction. In other words, the 
super operators analogous to those which are included in a construction of
the Lagrangians in a QFT do not appear in this consideration.
This indicates the difference in the schemes of the realization of the
supersymmetry ideas in nuclear and particle physics. 
The examples of the dynamical supersymmetries in nuclear structure
physics where the super operators have been explicitly introduced and
discussed are given in
\cite{Balantekin1,Morrison}.
However, in these examples the structure of supergroups is different
from that of QFT.

In this paper we consider a well known example of the
supersymmetry from the Interacting Boson--Fermion model, 
which on one side has a good experimental realization in $^{102}$Ru,
$^{103}$Rh and $^{104}$Pd \cite{Iachello1,Vervier}
and on the other side has an algebraic structure
which is similar to the SUSY model in QFT, however. It is the main aim of the
present paper to demonstrate this similarity and consider the consequences
of it.

\section{U(6/2) nuclear structure model}

Below we shall
assume that fermions occupy a single particle state with angular
momentum $j$=1/2 (e.g. a $p_{1/2}$ state). Thus, the underlying
graded algebra is $U(6/2)$. 

Before starting the consideration 
let us introduce a Hamiltonian
which has a simpler structure and which   can serve as an 
introduction to the more general case considered below.

The simplest Hamiltonian for a system of the noninteracting monopole 
$s$--bosons and a fermion with a spin equal $\frac{1}{2}$ and a
good helicity quantum number is 
\begin{eqnarray}
H=\epsilon (a^+_{\frac{1}{2}\frac{1}{2}}a_{\frac{1}{2}\frac{1}{2}} + s^+s)
\label{hamintr1_eq}
\end{eqnarray}
This is a standard example found in textbooks \cite{Kaku}, 
by which many features of the
supersymmetries can be illustrated.
However, this Hamiltonian does not possess rotational symmetry and 
is not realized in nuclear physics.
A simple extension of this Hamiltonian which possesses rotational symmetry is
\begin{eqnarray}
H=\epsilon(a^+_{\frac{1}{2}\frac{1}{2}}a_{\frac{1}{2}\frac{1}{2}} + a^+_{\frac{1}{2}-\frac{1}{2}}a_{\frac{1}{2}-\frac{1}{2}} + s^+s ).
\label{hamintr2_eq}
\end{eqnarray}
The super operators commuting with the Hamiltonian (\ref{hamintr2_eq}) are
$a^+_{\frac{1}{2}\frac{1}{2}}s$ and  $a^+_{\frac{1}{2}-\frac{1}{2}}s$. They
transform eigenstates of this Hamiltonian into another ones with the same
energy by changing the number of fermions in the state.

We go one step further and introduce the Hamiltonian which is 
used in the nuclear physics example
below and includes not only monopole
$s$--bosons but also quadrupole $d$--bosons.
We shall consider the case of the so
called $U(5)$ dynamical symmetry limit, for which the numbers of $s$-- and
$d$--bosons are conserved separately. In this case the 
Hamiltonian takes the form
\begin{eqnarray}
H=\epsilon (N_F + N_s) +H_d,
\label{ham_eq}
\end{eqnarray}
where $N_s \equiv s^+s$ and $N_F\equiv \sum_{m} a^+_{1/2m}a_{1/2m}$ 
are the number operators for the $s$--bosons
and the fermions, correspondingly, and
\begin{eqnarray}
H_d =\epsilon_d \sum_{\mu} d^+_{\mu} d_{\mu} + \sum_{L=0,2,4} C_L (d^+ d^+ )_{LM}(d d)_{LM}.
\label{hamd_eq}
\end{eqnarray}

The supercharges commuting with the Hamiltonian (\ref{ham_eq}) are
\begin{eqnarray}
P_{1/2 m} = a^+_{1/2 m} s,\hspace{1cm}   P^+_{1/2 m} =s^+ a_{1/2 m}, 
\label{super_eq}
\end{eqnarray}
\begin{eqnarray}
[ H, P_{1/2 m} ]=[ H, P^+_{1/2 m} ]= 0
\label{super1_eq}
\end{eqnarray}

Let us consider a graded algebra which contains $P_{1/2 m}$ and $P^+_{1/2 m}$
operators. 
The anticommutator of the  supercharge operators 
$P_{1/2 m}$ and  $P^+_{1/2 m}$  is
\begin{eqnarray}
\{ P_{1/2 m}, P^+_{1/2 m'}\} =\delta_{m m'} (N_s +\frac{1}{2} N_F )-\sqrt{2} (-1)^{1/2+m'} C^{1\eta}_{1/2m 1/2-m'} S_{1\eta},
\label{anticom_eq}
\end{eqnarray}
where $S_{1\eta}$
is the operator of the spin of the fermions
\begin{eqnarray}
S_{1\eta}=\frac{\sqrt{3}}{2}\sum_{m,m'}C^{1/2m'}_{1/2m 1\eta}a^+_{1/2m'}a_{1/2m},
\label{spindef_eq}
\end{eqnarray}
and $C^{a\alpha}_{b\beta  c\gamma}$ is a Clebsch--Gordan coefficient.
The next set of the anticommutators is simple 
\begin{eqnarray}
\{ P^+_{1/2 m}, P^+_{1/2 m'}\} = \{ P_{1/2 m}, P_{1/2 m'}\} =0,
\label{anticom2_eq}
\end{eqnarray}

The operators $N_s$ and $N_F$ commute with each other and with $S_{1\eta}$.
The operators $S_{1\eta}$ satisfy the usual commutation relations for
spin algebra
\begin{eqnarray}
\left [ S_{1\eta}, S_{1\eta '} \right ]=\sqrt{2} C^{1\eta ``}_{1\eta '  1\eta} S_{1\eta''}
\label{spin_eq}
\end{eqnarray}

Finally we consider the commutation relations of the operators
$N_s$, $N_F$ and $S_{1\eta}$
with the supercharges $P_{1/2m}$ and $P^+_{1/2m}$. They are
\begin{eqnarray}
\left [P_{1/2 m}, N_s \right ]= P_{1/2 m},
\label{super3a_eq}
\end{eqnarray}
\begin{eqnarray}
\left [P_{1/2 m}, N_F \right ]= -P_{1/2 m},
\label{super3b_eq}
\end{eqnarray}
\begin{eqnarray}
\left [P_{1/2 m}, S_{1\eta} \right ]= \frac{\sqrt{3}}{2} C^{1/2m'}_{1/2m  1\eta}P_{1/2 m'}.
\label{super3c_eq}
\end{eqnarray}

The operators $P_{1/2m}$, $P^+_{1/2m}$, $N_s$, $N_F$ and $S_{1\eta}$
form a graded algebra, which is $U(1/2)$, and the Hamiltonian (\ref{ham_eq})
corresponds to the following reduction chain
\begin{eqnarray}
U(6/2) \supset U(5)\otimes U(1/2)
\label{reduc_eq}
\end{eqnarray}

The multiplets of the eigenstates of the Hamiltonian (\ref{ham_eq})
with the same energy combine the states with different number of fermions.
For this reason they are called supermultiplets. Applying the super operators
$P_{1/2m}$, $P^+_{1/2m}$ we can transform one member of the supermultiplet
into an other one with different number of fermions.
All states belonging to the supermultiplet can be created 
by a repeated application of the supersymmetric
operators $P_{\frac{1}{2}\frac{1}{2}}$ and  $P_{\frac{1}{2}-\frac{1}{2}}$
to some basic state $|n_s , n_d >$ with $n_s$ and
$n_d$ being the numbers of the $s$-- and $d$-- bosons in this state, 
correspondingly. This state does not contain fermions. The complete 
supermultiplet thus includes the following states
\begin{eqnarray}
|n_s , n_d >\equiv (s^+ )^{n_s}\{(d^+)^{n_d}\}_{IM} |0>,\nonumber\\
P_{\frac{1}{2}\frac{1}{2}}|n_s ,n_d >, P_{\frac{1}{2}-\frac{1}{2}}|n_s ,n_d >, \nonumber\\
P_{\frac{1}{2}\frac{1}{2}}P_{\frac{1}{2}-\frac{1}{2}}|n_s ,n_d >.
\label{multiplet_eq}
\end{eqnarray}
Above the state $|n_s , n_d >$ is an analog of the Clifford (fermion) 
vacuum of the
Susy model of the QFT.
Examples of the complete supermultiplets are shown in Table 1.
Looking at these examples we see that there is an interesting relation
between the total number of the bosonic states, which include by
definition states without fermions and also states with even numbers
of fermions, and fermionic states, i.e. states with odd numbers of fermions,
in the supermultiplet. In the supermultiplets, which include the states with
the number of $s$--bosons equal or exceeding two, the total number of
the fermionic (magnetic) substates is equal to the number of the bosonic
(magnetic) substates.
In the supermultiplets, which include states with
the number of $s$--bosons equal or not exceeding one, the total number of
the fermionic (magnetic) substates is equal to twice the number of the bosonic
(magnetic) substates. This relation will be extended in the next section where 
the formal proof will be presented.

A nice and well known example of the approximate 
supersymmetry in nuclear physics corresponding to the
Hamiltonian (\ref{ham_eq}) is given in \cite{Iachello1,Vervier}. 
It is a complete supermultiplet containing the three nuclei
$^{102}_{44}$Ru$_{58}$ ($N_s$ + $N_d$=7),
$^{103}_{45}$Rh$_{58}$ ($N_s$ + $N_d$=6, $N_F$=1) and
also the two quasiparticle states of $^{104}_{46}$Pd$_{58}$
($N_s$ + $N_d$=5, $N_F$=2 ). However, the
experimental information on the two quasiparticle states 
in $^{104}$Pd is insufficient. 
A possible candidate for the lowest
state in $^{104}$Pd belonging to the supermultiplet is $0^+$ (1793 KeV)
state seen in $\beta ^-$ decay of $^{104}$Rh with $logft$=5.5. Experiments
on $^{103}$Rh ($^{3}$He,d) will be important to clarify a situation.
The odd proton
occupies the $p_{1/2}$ single particle state. The experimental 
spectra are given in
Fig.1. We put in Fig.1 the $0^+$ two quasiparticle state of $^{104}$Pd
at the same energy as the ground states of $^{102}$Ru and $^{103}$Rh
because in SUSY approach to nuclear structure we are dealing with the
energies of the states relative to the lowest state of the multiplet
in the same nucleus. This is also the reason
why in (\ref{ham_eq}) the same coefficient is used
in front of $N_F$ and $N_s$ terms. 
The corresponding 
full Hamiltonian $\bar{H}$
which includes total energies is
\begin{eqnarray}
\bar{H}=\epsilon_F N_F + \epsilon_s N_s + H_d.
\label{binding_eq}
\end{eqnarray}
If we consider only relative energies $\bar{H}$ is equivalent to $H$ of
(\ref{ham_eq}), however. This can be seen by rewriting $\bar{H}$
as follows
\begin{eqnarray}
\bar{H}=(\epsilon_F - \epsilon_s )N_F + \epsilon_s (N_F + N_s ) + H_d.
\label{binding1_eq}
\end{eqnarray}
One notes that the first term on the r.h.s. of (\ref{binding1_eq})
does not influence the relative energies. This is why this term is omitted
which leads to the Hamiltonian $H$ of (\ref{ham_eq}).

We see in Fig.1 that although the supersymmetry is 
realized with a good accuracy
in $^{102}$Ru and $^{103}$Rh
the symmetry is broken. So, to describe the experimental data in 
more details it is
necessary to add to the Hamiltonian (\ref{ham_eq}) a term which breaks the
supersymmetry described above. This term can be choosen as
\begin{eqnarray}
H_{break}= \alpha N_s\cdot N_F + \beta  \vec{L_d}\cdot \vec{S},
\label{break_eq}
\end{eqnarray}
where $\vec{L_d}=\sqrt{10}(d^+\tilde{d})_1$ is the operator of the orbital 
momentum of $d$--bosons.

\section{Formal comparison of supersymmetry in U(6/2) model 
and in quantum field theory}

In this section we shall compare the commutation and anticommutation
relations between the operators of the U(1/2) superalgebra considered
in the preceeding section with those of the SUSY model of the quantum field
theory. This comparison will demonstrate a close similarity between both
sets of relations and also some difference between them. The similarity
will be used further to derive a formal relation between the numbers 
of the bosonic and fermionic states in the supermultiplets. In QFT
the supercharges analogous to our
operators $P_{1/2 m}$ and  $P^+_{1/2 m}$ are denoted by $Q^L_{\alpha}$ and
$(Q^L_{\alpha})^+$, where $\alpha$ is a spinorial index and $L$ is an
index connected with an intrinsic symmetry group.
They commute with the 4--dimensional Lorentz momentum $P_{\mu}$
\begin{eqnarray}
\left [ Q^L_{\alpha} , P_{\mu} \right ] =\left [ (Q^L_{\alpha})^+ , P_{\mu} \right ]=0,
\label{momentum_eq}
\end{eqnarray}
and therefore with the Hamiltonian, which is a $\mu$=0 component 
of $P_{\mu}$.

The anticommutator of the supercharge operators $Q^L_{\alpha}$ and
$(Q^L_{\alpha})^+$ is
\begin{eqnarray}
\left \{Q^L_{\alpha}, (Q^M_{\beta})^+\right \}=2\delta^{LM}\sum_{\mu}\sigma^{\mu}_{\alpha \beta}P_{\mu},
\label{anticom1_eq}
\end{eqnarray}
where $\sigma ^{\mu}$ is a Dirac matrix.
The second anticommutator is
\begin{eqnarray}
\left \{Q^L_{\alpha}, Q^M_{\beta}\right \}=0.
\label{comparison3_eq}
\end{eqnarray}

In addition to the super operators
 $Q^L_{\alpha}$,
$(Q^L_{\alpha})^+$ and the Lorentz 4--dimensional momentum $P_{\mu}$
the SUSY model of QFT can includes also the boson type operators,
which form the intrinsic symmetry Lie algebra in QFT, where they 
are denoted by $B_l$. The operators $B_l$ satisfy to the following commutation relations.
\begin{eqnarray}
\left [B_l, B_m \right ]=i\sum_{k} f_{lmk}B_k,
\label{intrinsic_eq}
\end{eqnarray}
where $f_{lmk}$ are the structure constants.
The commutators between the intrinsic operators and super operators are
\begin{eqnarray}
\left [Q^L_{\alpha}, B_l\right ]=i\sum_{M} A^{LM}_l Q^M_{\alpha},
\label{qb_eq}
\end{eqnarray}

Now these relations can be putted in a correspondence with those 
relations obtained in the preceeding section. This is done below, where the 
corresponding relations are connected by the left--right arrows
\begin{eqnarray}
P_{1/2m}\Longleftrightarrow Q^L_{\alpha}, H \Longleftrightarrow P_{\mu}
\label{correspondencea_eq}
\end{eqnarray}
\begin{eqnarray}
N_s - N_F, S_{1\eta} \Longleftrightarrow B_l,
\label{correspondenceb_eq}
\end{eqnarray}
\begin{eqnarray}
\left [ P_{1/2 m}, H \right ]=0 \Longleftrightarrow \left [ Q^L_{\alpha}, P_{\mu} \right ]=0,
\label{correspondence2_eq}
\end{eqnarray}
\begin{eqnarray}
\left \{P_{1/2 m'}, P^{+}_{1/2 m}\right \}=\delta _{m m'}\left (N_s + \frac{1}{2} N_F\right )+\sqrt{3}\sum_{\eta}C^{1/2 m'}_{1/2 m  1\eta}S_{1 \eta}\nonumber\\
\Longleftrightarrow\left \{Q^L_{\alpha}, (Q^M_{\beta})^+\right \}=2\delta^{LM}\sum_{\mu}\sigma^{\mu}_{\alpha \beta}P_{\mu}
\label{correspondence3_eq}
\end{eqnarray}
\begin{eqnarray}
\left \{P_{1/2 m}, P_{1/2 m'}\right \}=0\Longleftrightarrow \left \{Q^L_{\alpha}, Q^M_{\beta}\right \}=0,
\label{correspondence4_eq}
\end{eqnarray}
\begin{eqnarray}
\left [ S_{1\eta}, S_{1\eta '} \right ]=\sqrt{2} C^{1\eta ''}_{1\eta '  1\eta} S_{1\eta ''} \Longleftrightarrow \left [B_l, B_m \right ]=i\sum_{k} f_{lmk}B_k
\label{correspondence5_eq}
\end{eqnarray}
\begin{eqnarray}
\left [P_{1/2 m}, N_s \right ]=P_{1/2 m}, \left [P_{1/2 m}, N_F \right ]=-P_{1/2 m},\nonumber\\ 
\left [ P_{1/2 m}, S_{1\eta}\right ]=-\frac{\sqrt{3}}{2}C^{1/2 m'}_{1/2 m  1\eta} P_{1/2 m'}
\Longleftrightarrow \left [Q^L_{\alpha}, B_l\right ]=i\sum_{M} A^{LM}_l Q^M_{\alpha}
\label{correspondence6_eq}
\end{eqnarray}
It is easy to see a strong similarity between analogous relations in the QFT
and in the nuclear case. This similarity is evident for the relations
(\ref{correspondence2_eq},\ref{correspondence4_eq}--\ref{correspondence6_eq}).
We notice, however, some difference in the structure of the expressions on
the right hand sides of the anticommutators in (\ref{correspondence3_eq}).
This distinction is due to the difference in the underlying bosonic algebras
in QFT and in nuclear case. This difference is reflected in the relation
between the numbers of the bosonic and fermionic states in the supermultiplets,
which we shall consider below.

As in quantum field theory this relation can be obtained
by multiplying
both sides of the relation (\ref{anticom_eq}) by $(-1)^{N_F}$ 
and taking a trace. 
On the left side we get
\begin{eqnarray}
Tr\left ((-1)^{N_F}P_{1/2m}P^+_{1/2m'}+(-1)^{N_F}P^+_{1/2m'}P_{1/2m}\right )=\nonumber\\Tr\left ((-1)^{N_F}P_{1/2m}P^+_{1/2m'}+P_{1/2m}(-1)^{N_F}P^+_{1/2m'}\right )=\nonumber\\Tr\left ((-1)^{N_F}P_{1/2m}P^+_{1/2m'}-(-1)^{N_F}P_{1/2m}P^+_{1/2m'}\right )=0
\label{trace_eq}
\end{eqnarray}
Above we have used the invariance of the trace operation under the cyclical 
permutations of the operators and the commutation relation (\ref{super3b_eq}).
Since the relation (\ref{anticom_eq}) can be also written as
\begin{eqnarray}
\{ P_{1/2m}, P^+_{1/2m'}\}=\delta_{m m'}s^+s +a^+_{1/2m}a_{1/2m'}
\label{anticom3_eq}
\end{eqnarray}
by taking $m=m'$ and summing over $m$ we get
\begin{eqnarray}
Tr\left((-1)^{\hat{N_F}}(2\hat{N_s} + \hat{N_F})\right )=0.
\label{rule1_eq}
\end{eqnarray}
In fact (\ref{rule1_eq}) contains the relation between the numbers of the 
bosonic and fermionic states in the supermultiplets.
To show it let us rewrite the trace in (\ref{rule1_eq}) in an explicit way 
as a sum of the diagonal matrix elements of the operator in circular
brackets. Since the super operators on the left of (\ref{anticom3_eq})
have nonzero matrix elements only between the states belonging to
the same supermultiplet this sum is restricted to these states
\begin{eqnarray}
\sum_{a\in supermultiplet}<a|(-1)^{\hat{N_F}}(2\hat{N_s}+\hat{N_F})|a>=0
\label{bosefermi1_eq}
\end{eqnarray}
Above the index $a$ includes bosonic states, i.e. the states without fermions
and with two fermions, and fermionic states, i.e. states with one fermion.
Since two fermions can be coupled in our case only to zero angular momentum
the number of magnetic substates without fermions is equal to the number
of magnetic substates with two fermions. Therefore, both are equal to the half
of the total number of the bosonic magnetic substates in the supermultiplet,
which we denote as $n_B$.
Let $n_F$ be the total number of the fermionic states in the
supermultiplet. Applying (\ref{bosefermi1_eq}) to the
supermultiplets with $N_s+N_F\equiv N^{max}_s\ge$ 2 one obtains
\begin{eqnarray}
\frac{1}{2}n_B\cdot 2N^{max}_s - n_F\left(2(N^{max}_s -1) + 1\right) +\frac{1}{2}n_B\left(2(N^{max}_s - 2)+2\right)\nonumber\\
=(n_B - n_F )(2N^{max}_s -1)=0
\label{bosefermi2_eq}
\end{eqnarray}
and thus $n_B =n_F$. Above $N^{max}_s$ is a maximum possible 
number of the $s$--bosons
in the states belonging to a supermultiplet.

For the supermultiplet with $N_s + N_F$=1 we get from (\ref{bosefermi1_eq})
\begin{eqnarray}
\frac{1}{2}n_B\cdot 1- n_F =0
\label{bosefermi3_eq}
\end{eqnarray}
Thus we can formulate the following rules:\\
--In the supermultiplets which include the states with
$N_s + N_F \ge 2$ 
the number of bosonic
(magnetic) substates is equal to the number of the fermionic 
(magnetic) substates.\\
--In the supermultiplets containing states with $N_s + N_F$ = 1 
the number of
the fermionic (magnetic) substates is equal to twice the number of the bosonic 
(magnetic) substates.\\
There is some difference in the relations between the numbers of  
bosonic and fermionic states in nuclear case and in QFT where for all
supermultiplets (excluding ground state) $n_F = n_B$. This difference is a
consequence of the difference in the anticommutation relations
(\ref{anticom_eq}) and (\ref{anticom1_eq}).

Since the relation between the numbers of the bosonic and fermionic 
(magnetic) substates
is a fundamental point in QFT, consider from this point of view 
the eigenstates of the 
Hamiltonians (\ref{hamintr1_eq},\ref{hamintr2_eq}). The supermultiplets of the
eigenstates of the Hamiltonian (\ref{hamintr1_eq}) correspond to the
supermultiplets of the QFT with the zero rest mass Clifford 
(fermion) vacuum. In this
case helicity is a good quantum number. The one--fermion state is, 
for instance, an analog of neutrino and a one $s$--boson state is 
an analog of S--neutrino.
In the case of the Hamiltonian (\ref{hamintr2_eq}) a one--fermion state is 
an analog of electron with two possible projections of the spin.
 
In the QFT the following expression has been derived for the Hamiltonian.
If we multiply both sides of (\ref{anticom1_eq}) by $\sigma^0_{\alpha \beta}$
and sum over $\alpha$ and $\beta$ we get
\begin{eqnarray}
H\equiv P_0 =\frac{1}{4}\sum_{\alpha , L}\left\{Q^L_{\alpha}, (Q^L_{\alpha})^+ \right\}.
\label{susyh1_eq}
\end{eqnarray}
In our case the relation, which can be derived in a similar way, is
different because of the difference in the anticommutation relations
(\ref{anticom_eq}) and (\ref{anticom1_eq})
\begin{eqnarray}
H = \frac{1}{2}N_F + \frac{1}{2}\sum_{m}\left\{P_{1/2m}, P^+_{1/2m}\right\} + H_d,
\label{susyh2_eq}
\end{eqnarray}
which is equivalent from the excitation energy point of view to
\begin{eqnarray}
H = \frac{1}{2}\sum_{m}\left\{P_{1/2m}, P^+_{1/2m}\right\} + H_d,
\label{susyh3_eq}
\end{eqnarray}
However, the ground state (i.e. the state with zero numbers of fermions,
$s$-- and $d$-- bosons) of our Hamiltonian has zero energy as in QFT.

\section{Conclusion}\indent

We have considered the comparison of the
supersymmetry concept in nuclear structure theory and in quantum
field theory. The nuclear structure model based on the U(6/2) graded
algebra is considered. It is shown that the U(6/2) algebra has a graded 
subalgebra whose generators commute with the model Hamiltonian. 
Thus, the existence
of the supersymmetric operators transforming bosons into fermions
and vice versa and commuting with the Hamiltonian is demonstrated.
We have furthermore obtained explicit expressions of the wavefunctions
of the various members of supermultiplets. For supermultiplets with
$N_F + N_s\ge 2$ we find an equal number of boson and fermion type
(magnetic) substates in the supermultiplet. This corresponds to the 
fundamental property of supermultiplets in QFT.

\acknowledgments
The authors would like to express their gratitude to
Profs. R.F.Casten, A.Gelberg, T.Otsuka, P. van Isacker and M.Zirnbauer
for discussions.
The work was supported in part by the DFG
under the contract Br 799/8-2.
One of the authors (R. V. J) is grateful to the
Universit\"at zu K\"oln for support.

\begin{table}
\caption
{Wave functions of the members of supermultiplets with various numbers of $s$-- and $d$--bosons. $P_{1/2m}=a^+_{1/2m}s$.}
\label{1_tab}
\begin{tabular}{cccc}
{\bf $N_s$ + $N_F$} & {\bf Boson} & {\bf Fermion} & {\bf Boson}\\
\tableline
 &  & 3 s--boson system &   \\
\tableline
3 & $s^+s^+s^+|0>$ & $P_{\frac{1}{2}\frac{1}{2}}s^+s^+s^+|0>=3 a^+_{\frac{1}{2}\frac{1}{2}}s^+s^+|0>$ & $ P_{\frac{1}{2}\frac{1}{2}} P_{\frac{1}{2}-\frac{1}{2}}s^+s^+s^+|0>=6a^+_{\frac{1}{2}\frac{1}{2}} a^+_{\frac{1}{2}-\frac{1}{2}}s^+|0>$\\
 &  &  $P_{\frac{1}{2}-\frac{1}{2}}s^+s^+s^+|0>=3 a^+_{\frac{1}{2}-\frac{1}{2}}s^+s^+|0>$ &     \\
\tableline
 & & 2 s--boson system &    \\
\tableline
 2 & $s^+ s^+|0>$ & $P_{\frac{1}{2}\frac{1}{2}}s^+ s^+ |0>=a^+_{\frac{1}{2}\frac{1}{2}}s^+ |0>$ & $P_{\frac{1}{2}\frac{1}{2}}P_{\frac{1}{2}-\frac{1}{2}}s^+ s^+ |0>=a^+_{\frac{1}{2}\frac{1}{2}}a^+_{\frac{1}{2}-\frac{1}{2}}|0>$\\
 &  & $P_{\frac{1}{2}-\frac{1}{2}}s^+ s^+ |0>=a^+_{\frac{1}{2}-\frac{1}{2}}s^+ |0>$ &   \\
\tableline
 & & 2s, 1d--boson system &  \\
\tableline
2 & $s^+s^+d^+_{\mu}|0>$ & $P_{\frac{1}{2}\frac{1}{2}}s^+s^+d^+_{\mu}|0>=2a^+_{\frac{1}{2}\frac{1}{2}}s^+d^+_{\mu}|0>$ & $P_{\frac{1}{2}\frac{1}{2}}P_{\frac{1}{2}-\frac{1}{2}}s^+s^+d^+_{\mu}|0>=2a^+_{\frac{1}{2}\frac{1}{2}}a^+_{\frac{1}{2}-\frac{1}{2}}d^+_{\mu}|0>$ \\
 &  & $P_{\frac{1}{2}-\frac{1}{2}}s^+s^+d^+_{\mu}|0>=2a^+_{\frac{1}{2}-\frac{1}{2}}s^+d^+_{\mu}|0>$ &    \\ 
\tableline
 &  & 1 s--boson system & \\
\tableline
1 & $s^+ |0>$ & $P_{\frac{1}{2}\frac{1}{2}} s^+ |0> = a^+_{\frac{1}{2}\frac{1}{2}} |0>$ & $P_{\frac{1}{2}\frac{1}{2}}P_{\frac{1}{2}-\frac{1}{2}} s^+ |0>$=0 \\
 & &  $P_{\frac{1}{2}-\frac{1}{2}} s^+ |0> = a^+_{\frac{1}{2}-\frac{1}{2}} |0>$ &  \\
\tableline
 &  & 1s,1d--boson system &   \\
\tableline
1 & $s^+d^+_{\mu}|0>$ & $P_{\frac{1}{2}\frac{1}{2}}s^+d^+_{\mu}|0>=a^+_{\frac{1}{2}\frac{1}{2}}d^+_{\mu}|0>$ &  $P_{\frac{1}{2}\frac{1}{2}}P_{\frac{1}{2}-\frac{1}{2}} s^+d^+_{\mu}|0>=0$    \\
 &  &  $P_{\frac{1}{2}-\frac{1}{2}}s^+d^+_{\mu}|0>=a^+_{\frac{1}{2}-\frac{1}{2}}d^+_{\mu}|0>$ &  \\ 
\tableline
  &  &  0 s--boson system &  \\
\tableline
0 & $|0>$ & $P_{\frac{1}{2}\frac{1}{2}}|0>$=0 &    \\
 & &  $P_{\frac{1}{2}-\frac{1}{2}}|0>$=0 &      \\
\end{tabular}
\end{table}

\begin{figure}
\centerline{\epsfig{file=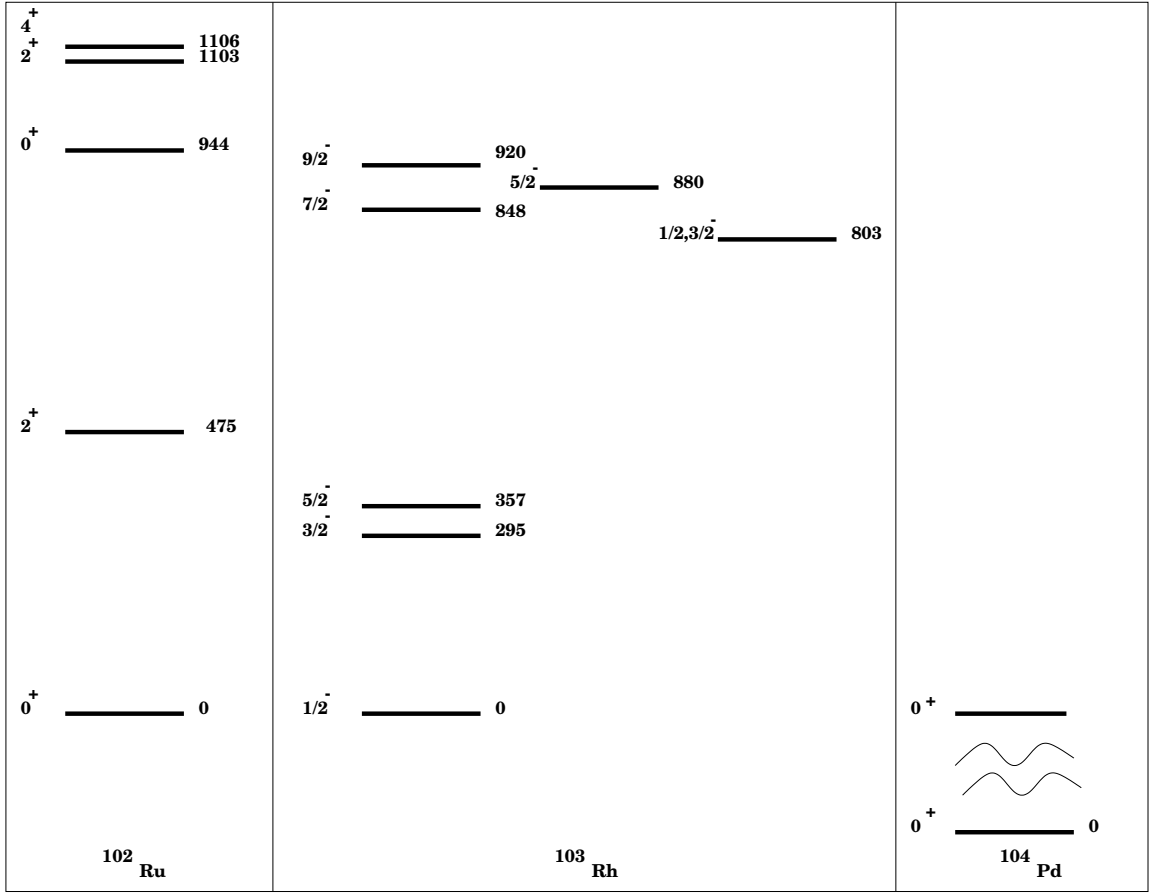,width=6in}}
\vspace*{10pt}
\caption{
An example of the U(6/2) supersymmetry in nuclei. A part of the experimental
spectra of the $^{102}$Ru--$^{103}$Rh--$^{104}$Pd supermultiplet
\protect\cite{Iachello1,Vervier} is shown. The energies are given in KeV.
The two quasiparticle states in $^{104}$Pd are not identified. 
A possible candidate for the lowest
state in $^{104}$Pd belonging to the supermultiplet is $0^+$(1793 KeV)
state seen in $\beta ^-$ decay of $^{104}$Rh with $logft$=5.5. Experiments
on $^{103}$Rh($^{3}$He,d) will be important to clarify a situation.
We note that
the number of boson type (magnetic) substates in the multiplet 
(in $^{102}$Ru and
$^{104}$Pd) is equal to the number of fermion type (magnetic) substates in
the multiplet (in $^{103}$Rh). This holds for some low lying multiplets 
as e.g. the multiplets
based on the $0^+$ and $2^+$ states in $^{102}$Ru.}
\label{1f_fig}
\end{figure}

\end{document}